\documentstyle[aps,preprint,pre,epsfig]{revtex}
\draft
\begin{document}

\title{Synchronization of Coupled Chaotic Dynamics on Networks}
\author {R. E. Amritkar \footnote{e-mail: amritkar@prl.errnet.in} and
Sarika Jalan\footnote{e-mail: sarika@prl.ernet.in}}
\address{Physical Research Laboratory, Navrangpura, Ahmedabad 380 009, India.}
\maketitle

\begin{abstract}
We review some recent work on the synchronization of coupled dynamical 
systems on a variety of networks. 
When nodes show synchronized behaviour, two interesting 
phenomena can be observed. First,
 there are some nodes
of the floating type that show intermittent behaviour between getting
attached to some clusters and evolving independently. Secondly, two
different ways of cluster formation can be identified, 
namely self-organized clusters which have mostly
intra-cluster couplings and driven clusters which have mostly 
inter-cluster couplings. 
\end{abstract}

\section{Introduction}

The phenomena of synchronization of several dynamical variables or
oscillators is
important for many physical and biological systems \cite{syn-book}. Some
important physical systems showing synchronization are arrays of
lasers \cite{syn-laser}, microwave oscillators and
superconducting Josephson junctions \cite{syn-book} while some
important biological systems are synchronous firing of neurons
\cite{syn-neuron},
networks of pacemaker cells in the heart \cite{syn-heart}, metabolic
synchrony in yeast cell suspensions \cite{syn-meta}, congregations of 
synchronously flashing fireflies \cite{syn-fire}, and cricket that chirp
in unison \cite{syn-cric}.

Coupled oscillators were first studied by Winfree \cite{winfree} and
Kuramoto \cite{kuramoto}. The Kuramoto model describes a large
population of coupled limit cycle oscillators with random
frequencies. If the coupling strength exceeds a critical threshold, the 
system exhibits a phase transition to a synchronous state where
several oscillators synchronize and lock to a common frequency \cite{winfree1}.

Two important developments have regenerated interest in the study of
synchronization of dynamical variables. First is the recognition that
chaotic dynamical systems can show exact or phase
synchronization \cite{chaos-syn-pec-car}. Second is the observation that
several natural
systems have an underlying geometric structure which can be described
by complex networks \cite{net-rev-strogatz,net-rev-barabasi}. This has 
opened up the possibility of discovering new 
interesting phenomena in coupled dynamical systems on complex
networks. We discuss some of these aspects in this article.

Several networks in the real world consist of dynamical elements interacting
with each other and the interactions define the links of the network.
Several of these networks have a large number
of degrees of freedom and it is important to understand their dynamical
behaviour. A general model of coupled dynamical systems on
networks will consist
of the following three elements.
\begin{enumerate}
\item The evolution of uncoupled elements.
\item The nature of couplings.
\item The topology of the network.
\end{enumerate}  

Most of the earlier studies of synchronized cluster formation in
coupled chaotic systems have focused on
networks with large number of connections ($\sim N^2$)
\cite{rev-Kaneko} or nearest neighbour couplings on lattice models. 
Recently, we have considered complex networks
with number of connections
of the order of $N$ \cite{sarika-REA1}. This small number of connections
allows us to study the role that different connections play in
synchronizing different nodes and
the mechanism of synchronized cluster formation. The study reveals two 
interesting phenomena. First,
when nodes form synchronized clusters, there can be some nodes which
show an intermittent behaviour between independent evolution and
evolution synchronized with some cluster. Secondly,  
the cluster formation can be in two different ways, driven and self-organized
phase synchronization \cite{sarika-REA1}.
The connections or couplings in the self-organized phase synchronized
clusters are mostly of the intra-cluster type while those in the
driven phase synchronized clusters are mostly of
the inter-cluster type. We will briefly review these features in this article.

\section{Synchronization of dynamical systems}

Synchronization of different dynamical variables can be defined in
several ways. Exact synchronization corresponds to the situation where 
the dynamical
variables have identical values, i.e. two dynamical variables
$x_i$ and $x_j$ are 
exactly synchronized if $x_i(t) = x_j(t)$
\cite{chaos-coup1}. Generalised synchronization 
is defined by some functional relation between the dynamical
variables \cite{generalised}. Several other types of synchronization
such as phase
synchronization \cite{phase1}, lag synchronization \cite{lag}, 
anticipatory synchronization \cite{anticipatory} etc. have been defined.
 
\subsection{Kuramoto model of coupled oscillators}

Kuramoto model of coupled oscillators can be introduced through the evolution
equation \cite{kuramoto}
\begin{equation}
\dot\theta_i = \omega_i + \frac{K}{N} \sum_{j=1}^N \sin(\theta_j -
\theta_i), \, \, i=1,\ldots,N,
\label{Kuramoto}
\end{equation}
where $\theta_i$ is the phase of oscillator $i$, $\omega_i$ is its natural 
frequency and $K$ is the coupling strength. Phase
synchronization of the oscillators can be studied using the complex order
parameter defined as 
\[ r e^{i \psi} = \frac{1}{N} \sum_{j=1}^N e^{i \theta_j} \]
 where $r$ 
measures the phase coherence, and $\psi$ is the average
phase. Clearly, $r$ vanishes for a uniform distribution of phases
$\theta_i$, and $r=1$ if all $\theta_i$ are equal. The evolution
equation for the phases can be written as,
\[ \dot\theta_i = \omega_i + K r sin(\psi - \theta_i)\]
This is a mean field form of the equation.
For $K$ less than some
threshold $K_c$, the oscillators
show unsynchronized behaviour. But, when $K > K_c$,
the oscillator population splits into two groups, 
the oscillators near the center of the
frequency distribution lock together at some frequency $\omega_0$ to
form a synchronized cluster, while 
those in the tail retain their natural frequencies and drift relative to
the synchronized cluster. This state is often called partially synchronized 
state. With further increase in $K$, more and more oscillators are
recruited into the synchronized clusters 
\cite{kuramoto,crawford,rev-CO-strogatz}. 

Kuramoto's model was originally motivated
by biological oscillators \cite{winfree}, but it has found applications in
many diverse systems 
such as flavour evolution of neutrinos \cite{osc-neutrino},
arrays of Josephson junction \cite{syn-josephson}, semiconductor lasers
\cite{syn-laser1} and in several other systems 
\cite{kuramoto-ex1,kuramoto-ex2}.

\subsection{Synchronization of chaotic systems}

Pecora and Carroll showed that two identical chaotic systems can
synchronize if approrpiate driving mechanisms are introduced
\cite{chaos-syn-pec-car}. 
Let ${\bf x}$ be an $n$ dimensional dynamical variable evolving as
(drive system)
\begin{equation}
\dot{\bf x} = {\bf f}({\bf x}).
\label{dyn-nd}
\end{equation}
Divide the dynamical variables into two parts, 
${\bf x} = ({\bf x}_d,{\bf x}_r)$, a drive part and a response part. 
Consider another dynamical system (response system) given by
\begin{equation}
\dot{\bf x}^{'} = {\bf f}({\bf x}_d,{\bf x}_r^{'}).
\label{response}
\end{equation} 
where the drive variables ${\bf x_d}$ are obtained from Eq.~(\ref{dyn-nd}).
Under suitable conditions, i.e. the conditional Lyapunov exponents are 
negative, the response variable ${\bf x}_r^{'}$ synchronize with those
of the drive system, ${\bf x}_r$. The conditional Lyapunov exponents
are obtained by considering the subspace of response variables. The
important interesting feature is that the variables synchronize even
when they are evolving chaotically. Thus we do not have frequency
locking, but we can have phase synchronization if a suitable phase
variable can be defined \cite{phase1}. Such phase synchronization is
observed in several biological systems \cite{syn-book}

\subsection{Coupled dynamics on Complex Networks}

Several complex systems have underlying structures that are 
described by networks or graphs
\cite{net-rev-strogatz,net-rev-barabasi}. Recent interest in networks is due to
the discovery that several naturally occurring
networks come under some universal classes and they can
be simulated with simple mathematical models, viz small-world 
networks \cite{smallworld-watt}, scale-free networks \cite{scalefree} etc. 

Consider a network of $N$ nodes and $N_c$ connections (or couplings)
between the nodes. Let each node of the network be assigned an $m$-dimensional
dynamical variable ${\bf x}^i, i=1,2,\ldots,N$. A very general dynamical
evolution can be written as
\begin{equation}
\frac{d{\bf x}_i}{dt} = {\bf F}(\{ {\bf x}_i \}).
\end{equation}
Here, we consider a separable case and the evolution equation 
can be written as,
\begin{equation}
\frac{d{\bf x}_i}{dt} = {\bf f}({\bf x}_i) + \frac{\epsilon}{k_i}
\sum_{j \in \{k_i\}} {\bf g}({\bf x}_j).
\label{evol-cont}
\end{equation}
where $\epsilon$ is the coupling constant, $k_i$ is the degree of node
$i$, and $\{k_i\}$ is the set of nodes connected to the node $i$.
A sort of diffusion version of the evolution equation~(\ref{evol-cont}) is
\begin{equation}
\frac{d{\bf x}_i}{dt} = {\bf f}({\bf x}_i) + \frac{\epsilon}{k_i}
\sum_{j \in \{k_i\}} \left({\bf g}({\bf x}_j)-{\bf g}({\bf
x}_i)\right).
\label{evol-cont-diff}
\end{equation} 

Discrete versions of Eqs.~(\ref{evol-cont})
and~(\ref{evol-cont-diff}) are
\begin{equation}
{\bf x}_i(t+1) = {\bf f}({\bf x}_i(t)) + \frac{\epsilon}{k_i}
\sum_{j \in \{k_i\}} {\bf g}({\bf x}_j(t)).
\label{evol-disc}
\end{equation}
and
\begin{equation}
{\bf x}_i(t+1) = {\bf f}({\bf x}_i(t)) + \frac{\epsilon}{k_i}
\sum_{j \in \{k_i\}} \left({\bf g}({\bf x}_j(t))-{\bf g}({\bf
x}_i(t))\right).
\label{evol-disc-diff}
\end{equation}  

In numerical studies, for the discrete evolution we can use any 
descrete map such as
logistic or circle maps while for 
the continuous case we can use chaotic systems such as Lorenz or 
R\"ossler systems.

As noted before synchronization of coupled dynamical systems 
\cite{syn-book} is manifested by the appearance of some
relation between the functionals of different dynamical variables. 
When the number 
of connections in the network
is small ($N_C \sim N$) and when the local dynamics of the
nodes (i.e. function $f(x)$) is in the chaotic zone, and we look at
exact synchronization, only
few synchronized clusters with small number of nodes are formed.
However, when we look at phase synchronization, synchronized clusters 
with larger number of nodes are obtained. 

\section{General properties of synchronized dynamics on complex networks}

We consider some general properties of synchronized dynamics. 
They are valid
for any coupled discrete and continuous dynamical systems.
Also, these
properties are applicable for exact as well as phase or any other type 
of synchronization
and are independent of the type of network. 

\subsection{Behavior of individual nodes}
As the network evolves, it splits into several synchronized clusters. 
Depending on their asymptotic dynamical behaviour the nodes
of the network can be divided into three types. \\
(a) {\it Cluster nodes}: A node of this type synchronizes with other nodes and 
forms a synchronized cluster. Once this node enters a synchronized cluster
it remains in that cluster afterwards. \\
(b) {\it Isolated nodes}: A node of this type does not synchronize
with any other node 
and remains isolated for all the time. \\
(c) {\it Floating Nodes}:  A node of this type keeps on switching intermittently
between an independent evolution and a synchronized evolution
attached to some cluster.

Of particular interest are the floating nodes and we will discuss some
of their properties afterwards.

\subsection{Mechanism of cluster formation} 
The study of the relation between the synchronized clusters and the couplings
between the nodes represented by the
adjacency matrix $C$ shows two different 
mechanisms of cluster formation \cite{sarika-REA1,pre2}. \\
(i) Self-organized clusters: The nodes of a cluster can be 
synchronized because of intra-cluster
couplings. We refer to this as
the self-organized synchronization and 
the corresponding synchronized clusters as self-organized clusters. \\
(ii) Driven clusters: The nodes of a cluster can be
synchronized because
of inter-cluster couplings. 
Here the nodes of one cluster are driven by those
of the others. We refer to this as the driven
synchronization and the corresponding clusters as driven clusters.

\begin{figure}[hb]
\begin{center}
\includegraphics[width=14cm]{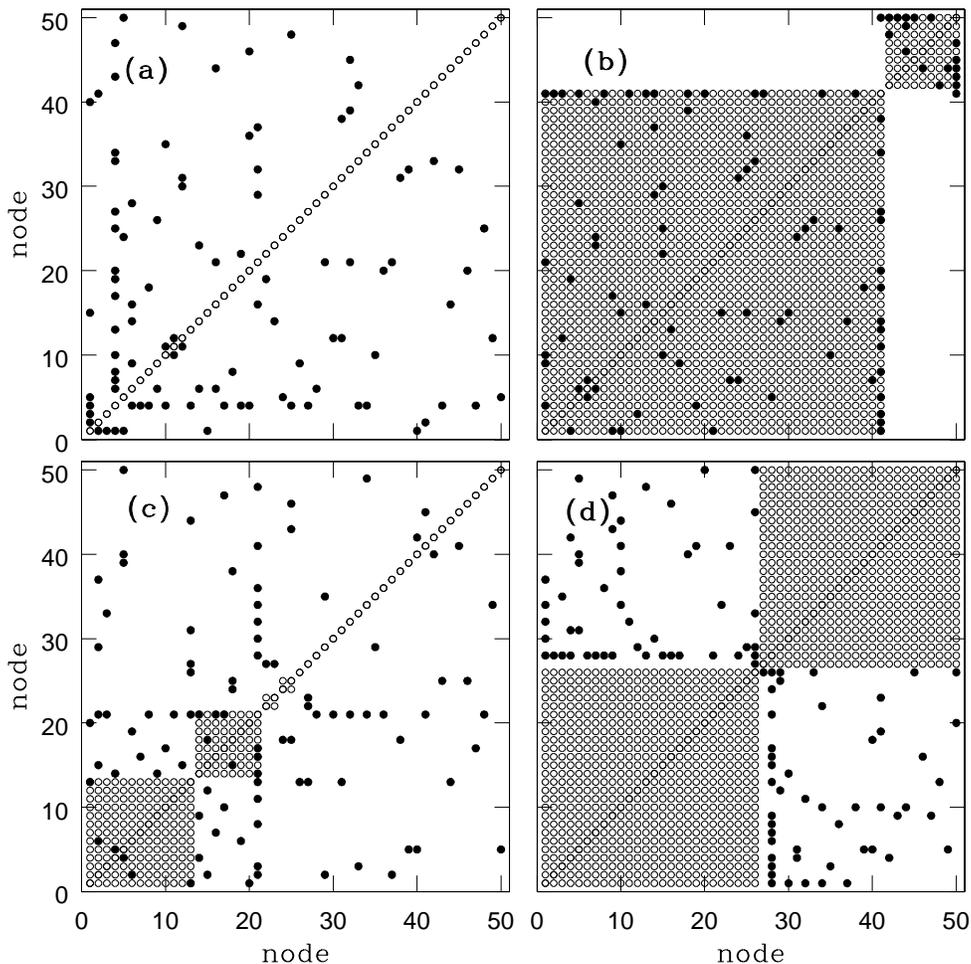}
\end{center}
\caption{The figure shows node versus node diagrams for several
examples illustrating the  
self-organized and driven phase synchronization and the
variety of clusters that
are formed. The examples are for coupled maps on a scale free network with
$N=50$, the local dynamics is given by the logistic $4x(1-x)$ and 
the coupling function is linear. The solid
circles show that the two corresponding nodes are coupled (i.e. $C_{ij}=1$) 
and the open
circles show that the corresponding nodes are phase
synchronized. In each case the node numbers are reorganized so that
nodes belonging to the same cluster are 
numbered consecutively and the clusters get displayed in decreasing
sizes. (a) Figure shows turbulent phase
for coupling constant $\epsilon=0.10$.
(b) An ideal self-organized phase synchronization for
$\epsilon=0.16$. (c) Mixed behavior for
$\epsilon=0.32$. (d) An ideal driven phase synchronization
for $\epsilon=0.90$.}
\label{scale-x-clus}
\end{figure}

In numerical studies it is possible to observe ideal clusters
of both the types,
as well as clusters of the mixed type where both ways of
synchronization contribute
to cluster formation. Fig.~\ref{scale-x-clus} shows some examples of
ideal as well as
mixed clusters in coupled map networks \cite{sarika-REA1}. 
In general we find
that the scale free
networks and the Caley tree networks lead to better cluster formation than 
the other types of networks with the same average connectivity. 

\begin{figure}[hb]
\begin{center}
\includegraphics[width=14cm]{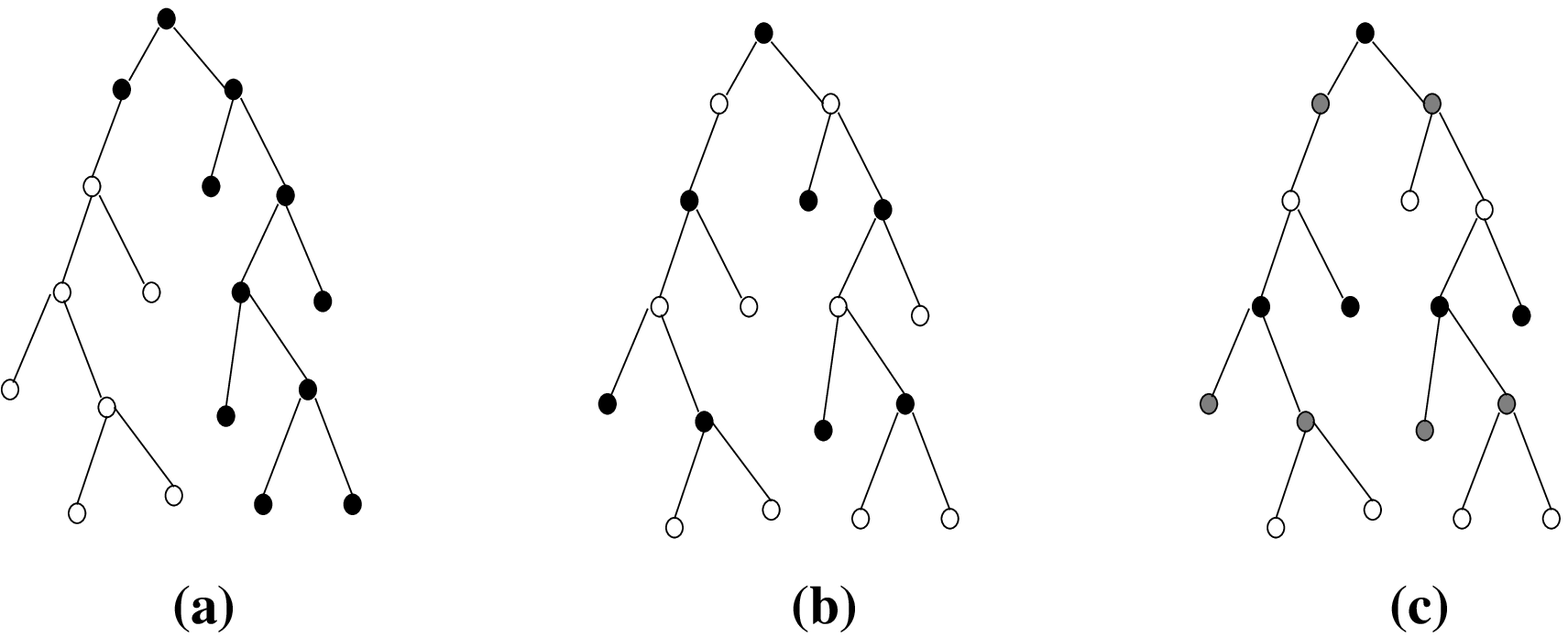}
\end{center}
\caption{Different ways of cluster formation in a tree structure are
demonstrated. The open, solid and gray circles show nodes belonging to 
different clusters. (a) shows two clusters of the self-organized
type, (b) shows two clusters of driven type and (c) shows three
clusters of the driven type.}
\label{tree-clus}
\end{figure}

Geometrically the two mechanisms of cluster formation can be easily
understood by considering a tree type network. A tree
can be broken into different clusters in different ways. \\
(a) A tree can be broken into two or more disjoint clusters with only
intra-cluster couplings by breaking one
or more connections. Clearly, this splitting is not unique
and will lead to self-organized clusters. Figure~\ref{tree-clus}(a)
shows a tree forming two synchronized clusters of self-organized
type. This situation is
similar to an Ising ferromagnet where domains of up and down spins can 
be formed. \\
(b) A tree can 
also be divided into two clusters by putting connected nodes into different
clusters. This division is unique and leads to two clusters with only
inter-cluster couplings, i.e. driven clusters. Figure~\ref{tree-clus}(b)
shows a tree forming two synchronized clusters of the driven type. 
This situation is
similar to an Ising anti-ferromagnet where two sub-lattices of up and 
down spins are formed.\\
(c) Several other ways of splitting a tree are possible. E.g. it is easy
to see that a tree can be broken into three clusters of the driven
type. This is shown in figure~\ref{tree-clus}(c). There is no simple
magnetic analog for this type of cluster formation. It
can be observed close to a period
three orbit. We note that four or more clusters of the driven type are 
also possible. As compared to the cases (a) and (b) discussed above
which are commonly observed, 
the clusters of case (c) are not so common and are
observed only for some values of the parameters. 

\section{Linear stability analysis}

A suitable network to study the stability of self-organized
synchronized clusters is the globally coupled network. The stability
of globally coupled maps is well studied in the literature
\cite{GCM-stab1,GCM-stab2,GCM-stab3}.
An ideal example to consider the stability of the driven synchronized
state is a complete bipartite network.
A complete bipartite network consists of two sets of nodes with each
node of one set connected
with all the nodes of the other set and no connection between the
nodes of the same set. Let $N_1$ and $N_2$ be the number of nodes 
belonging to
the two sets. We define a bipartite synchronized
state as the one that 
has all $N_1$ elements of the first set synchronized
to some value, say ${\bf X}_1(t)$, and 
all $N_2$ elements of the second set  synchronized
to some other value, say ${\bf X}_2(t)$.

 All the eigenvectors and the eigenvalues of the Jacobian matrix for the bipartite synchronized state can be determined explicitly.
The eigenvectors of the type
$(\alpha,\ldots,\alpha,\beta,\ldots,\beta)^T$ determine the
synchronization manifold and this manifold has dimension two. 
All other eigenvectors correspond to
the transverse manifold.
Lyapunov exponents corresponding to the transverse eigenvectors
for Eq.~(\ref{evol-disc-diff}) with one dimensional variables and $g(x)=f(x)$
are
\begin{eqnarray}
\lambda_1 &=& \ln|(1-\epsilon)| + \frac{1}{\tau} \lim_{\tau \to\infty}
\sum^\tau_{t=1} \ln |f^{\prime}(X_1)|, 
\nonumber \\
\lambda_2 &=& \ln|(1-\epsilon)| + \frac{1}{\tau} \lim_{\tau \to\infty}
\sum^\tau_{t=1} \ln |f^{\prime}(X_2)|,
\label{lya-driven-Nlarge}
\end{eqnarray}
and $\lambda_1$ and $\lambda_2$ are respectively $N_1-1$ and $N_2-1$
fold degenerate \cite{pre2}. Here, $f^{\prime}_(X_1)$ and $f^{\prime}_(X_2)$ are
the derivatives of $f(x)$  at $X_1$ and $X_2$ respectively.
The synchronized state is stable provided the transverse Lyapunov
exponents are negative.
If $f^\prime$ is bounded then from
Eqs.~(\ref{lya-driven-Nlarge}) we see that for $\epsilon$ larger than
some critical value, $\epsilon_b (<1)$, bipartite synchronized state will be
stable. Note that this bipartite synchronized state will be stable even if one
or both the remaining Lyapunov exponents corresponding to the
synchronization manifold are
positive, i.e. the trajectories are chaotic. The linear stability 
analysis for other type of couplings and dynamical systems can be done 
along similar lines.

\section{Floating nodes}
We had noted earlier that the nodes can be divided into three types,
namely cluster nodes, isolated nodes and floating nodes, depending on
the asymptotic behavior of the nodes. 
Here, we discuss some properties of the floating nodes which show an
intermittent behavior between synchronized evolution with some cluster 
and an independent evolution.

\begin{figure}
\begin{center}
\includegraphics[width=10cm]{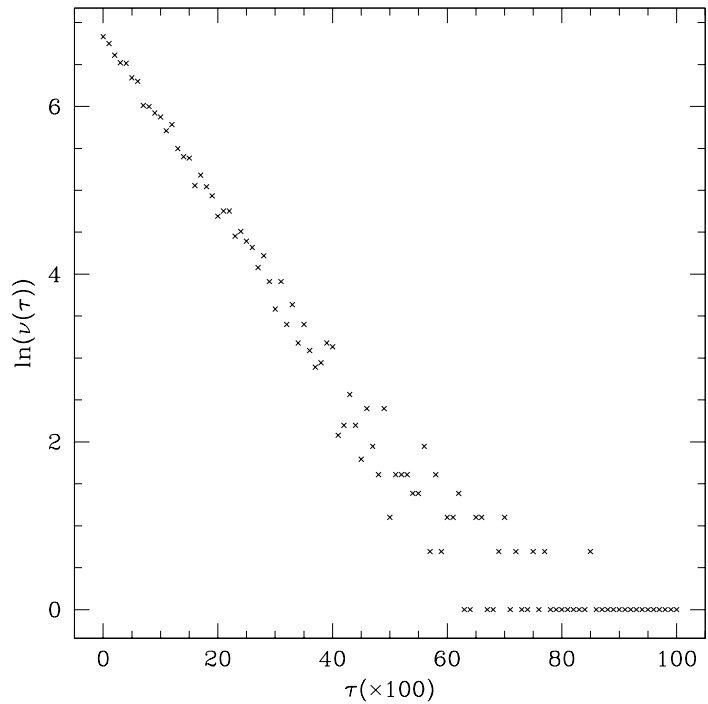}
\end{center}
\caption{The figure plots the frequency of residence time $\nu(\tau)$
of a floating node in a cluster as a
function of the residence time $\tau$. A good straight line fit on log-linear
plot shows exponential dependence.}
\label{freq-floating}
\end{figure}

Let $\tau$ denote the residence time of a floating node in a cluster 
(i.e. the continuous time interval that the node is in a cluster).
Figure~\ref{freq-floating} plots the frequency of residence time $\nu(\tau)$
of a floating node as a
function of the residence time $\tau$. 
A good straight line fit on log-linear
plot shows an exponential dependence, 
\begin{equation}
\nu(\tau) \sim \exp(-\tau/ \tau_r)
\label{exp-dist-floating}
\end{equation} 
where
$\tau_r$ is the typical residence time
for a given node. We have also studied the distribution of the time
intervals for which a floating node is not synchronized with a given
cluster. This also shows an exponential distribution.  

Several natural systems show examples of floating nodes, 
e.g. some birds may show intermittent behaviour between free flying and
flying in a flock. An interesting example in physics is that of particles
or molecules in a liquid in
equilibrium with its vapor where the particles
intermittently belong to the liquid and vapor. Under suitable
conditions it is possible to 
argue that the residence time of a tagged particle in the liquid phase 
should have an exponential distribution \cite{pre2}, i.e. a behavior similar to
that of the floating nodes (Eq.~(\ref{exp-dist-floating})).

\section{Conclusion and Discussion}
We have briefly discussed synchronization properties of
coupled dynamical elements on
complex networks. In the course of time evolution these dynamical
elements form synchronized clusters. 

In several cases when synchronized clusters are formed there are some
isolated nodes which do not belong to any cluster. More interestingly
there are some floating nodes which show an intermittent behavior
between an independent evolution and an evolution synchronized with some
cluster. The residence time spent by a floating node in the synchronized cluster 
shows an exponential distribution.

We have identified two mechanisms of cluster formation, 
self-organized and driven phase synchronization. For self-organized
clusters intra-cluster couplings dominate while for driven clusters
inter-cluster couplings dominate.

\end{document}